\documentclass[natbib,conference]{IEEEtran}
\IEEEoverridecommandlockouts
\usepackage{multirow}
\usepackage{cite}
\usepackage{amsmath,amssymb,amsfonts}
\usepackage{algorithmic}
\usepackage{graphicx}
\usepackage{textcomp}
\usepackage{xcolor}

\usepackage[colorlinks = true,
            linkcolor = blue,
            urlcolor  = blue,
            citecolor = blue,
            anchorcolor = blue]{hyperref}

\begin{document}

\title{ABSP System for The Third DIHARD Challenge}

\author{\IEEEauthorblockN{A Kishore Kumar$^1$, Shefali Waldekar$^1$, Goutam Saha$^1$, Md Sahidullah$^2$}
\IEEEauthorblockA{$^1$Dept. of Electronics and ECE, Indian Institute of Technology Kharagpur, Kharagpur, India\\
$^2$Universit\'{e} de Lorraine, CNRS, Inria, LORIA, F-54000, Nancy, France}
\url{kishore@iitkgp.ac.in}}

\maketitle

\begin{abstract}
This report describes the speaker diarization system developed by the ABSP Laboratory team for the third DIHARD speech diarization challenge. Our primary contribution is to develop acoustic domain identification (ADI) system for speaker diarization. We investigate speaker embeddings based ADI system. We apply a domain-dependent threshold for agglomerative hierarchical clustering. Besides, we optimize the parameters for PCA-based dimensionality reduction in a domain-dependent way. Our method of integrating domain-based processing schemes in the baseline system of the challenge achieved a relative improvement of $9.63\%$ and $10.64\%$ in DER for core and full conditions, respectively, for Track 1 of the DIHARD III evaluation set.

\end{abstract}


\section{Notable highlights}
We participated in the \textbf{Track 1} of the third DIHARD challenge~\cite{ryant2020third}. Our main focus was to apply domain-dependent processing which was found promising in preliminary studies with the second DIHARD dataset~\cite{sahidullah2019speed, fennir2020acoustic}. We propose a simple modification of the baseline system of the challenge which results considerable reduction of the error rates compared to the baseline performance. The notable features of our submission to the challenge are as follows:

\begin{itemize}
    
    \item We propose a simple but efficient method for acoustic domain identification (ADI) using speaker embeddings of the full-recording. We observed that i-vector-based speaker embeddings are considerably better than x-vector-based speaker embeddings for ADI task. 
    
    \item We have found that that full domain-dependent processing with domain-dependent clustering and domain-dependent probabilistic linear discriminant analysis (PLDA) adaptation does not improve the diarization performance. However, this helps when the clustering is done in a domain-dependent way, but PLDA adaptation during scoring is made with audio-data from all the eleven domains.
    
    \item We also found that experimental optimization of the parameters for principal component analysis (PCA) in a domain-specific way further improves the diarization performance.
    
    \item The proposed system does not introduce much computational overhead over the baseline system for the diarization. Though this approach requires more time for empirical optimization of the parameters on the development set, the additional computational cost is negligible for the evaluation data.
    
    \item The proposed system does not have any fusion or system combination from evaluation perspective. Considering the fact that most of the top systems in this challenge are combination of two or more sub-systems, our algorithm is remarkably faster than other competitive systems.  
    
\end{itemize}

\section{Data resources}
The ABSP system has two major components: ADI and speaker diarization. The ADI system uses i-vector speaker embeddings extracted with models trained on VoxCeleb 1\footnote{\url{https://www.robots.ox.ac.uk/~vgg/data/voxceleb/vox1.html}} and 2\footnote{\url{https://www.robots.ox.ac.uk/~vgg/data/voxceleb/vox2.html}} corpora.

On the other hand, the diarization system uses an embedding extractor trained on a combination of VoxCeleb 1 and VoxCeleb 2 augmented with additive noise and reverberation from MUSAN\footnote{\url{https://www.openslr.org/17/}} and RIR\footnote{\url{https://www.openslr.org/28/}} database, respectively.

\section{Detailed description of algorithm}
Our diarization system is primarily based on the baseline system created by the organizers~\cite{ryant2020thirdpaper}. We have used the toolkit\footnote{\url{https://github.com/dihardchallenge/dihard3_baseline}} with the same frame-level acoustic features, embedding extractor, scoring method, etc. The ADI system is based on the speaker embeddings as sentence-level feature and nearest neighbor classifier~\cite{kumar2021domain}. In order to extract utterance-level embeddings for ADI task, we used pre-trained i-vector~\cite{dehak2010front} model trained on VoxCeleb audio-data\footnote{\url{https://kaldi-asr.org/models/m7}}.

We can summarize the steps for the speaker diarization as follows:

\begin{enumerate}
    \item \emph{ADI task}: First, the ADI system was developed from the development set. We have used nearest neighbour classifier with cosine similarity. The full development set with all 254 files was used for training the final ADI system. More details about this system are reported in~\cite{kumar2021domain}.
    
    \item \emph{Domain-dependent threshold selection}: The baseline system for the challenge finds the optimum threshold by computing diarization error rates (DERs) on full development set at different thresholds ranging from $-1.5$ to $0.0$\footnote{\url{https://github.com/dihardchallenge/dihard3_baseline/blob/master/recipes/track1/local/diarize.sh}}. We follow the same process but for different acoustic-domains, independently. At the end of this step, the optimum thresholds for each domain are stored in a lookup table. 
    
    \item \emph{Domain-dependent dimensionality reduction}: The PLDA scoring involves dimensionality reduction of the embedding using PCA. The baseline system preserves 30\% of the total energy during dimensionality reduction. Instead of applying fixed value of $0.3$ for all the recording, we optimized this for each domain separately by varying it between $0.1$ to $0.9$ with a step of $0.1$. Similar to the previous step, the optimum parameters for each domain are preserved in another lookup table.     
    
    \item \emph{Diarization on the evaluation set}: Finally, during the diarization on the evaluation set, we first computed the i-vector of the full-recording to the be processed. Then, we predicted the corresponding acoustic domain using the ADI system. This is followed by the selection of clustering threshold and dimensionality reduction parameters corresponding to the predicted labels. 
    
\end{enumerate}

\section{Results on the development set}
The speaker diarization results on development set are shown in Table~\ref{Table:Dev}. We have also shown the results for the evaluaiton set in Table~\ref{Table:Eval}. Both these results confirm considerable improvement over baseline system.

\begin{table}[h]
  \caption{Results showing the speaker diarization performance using baseline and proposed methods on \textbf{development set}.}

  \begin{center}
  \label{Table:Dev}
  \centering
  \renewcommand{\arraystretch}{1.2}
  \begin{tabular}{|c|c|c|c|c|}
    \hline
    \multirow{2}{*}{Method} & \multicolumn{2}{|c|}{Full} & \multicolumn{2}{|c|}{Core}\\
     \cline{2-5}
      & DER & JER & DER & JER \\
    \hline
    Baseline	& 19.59	&43.01	&20.17&	47.28\\
    Proposed	& 17.40&	38.08&	17.95&	42.12\\
    \hline
  \end{tabular}

  \end{center}

\end{table}

\begin{table}[h]
  \caption{Same as Table ~\ref{Table:Dev} but for \textbf{evaluation set}.}
  \begin{center}
  \label{Table:Eval}
  \centering
  \renewcommand{\arraystretch}{1.2}
  \begin{tabular}{|c|c|c|c|c|}
    \hline
    \multirow{2}{*}{Method} & \multicolumn{2}{|c|}{Full} & \multicolumn{2}{|c|}{Core}\\
     \cline{2-5}
       & DER & JER & DER & JER \\
    \hline
    Baseline 	& 19.19	&43.28&	20.39&	48.61\\
    Proposed 	& 17.20	&37.30&	18.66&	42.23\\
    \hline
  \end{tabular}
  \end{center}
\end{table}

\section{Hardware requirements}
The codes were run on Dell PowerEdge R730 server\footnote{\url{https://www.dell.com/en-us/work/shop/povw/poweredge-r730}}. It has Intel Xeon (R) CPU E5-2695 v4 @ 2.10GHz processor and 128 GB memory with 72 CPUs, 2 sockets, 18 cores/socket, and 2 threads/core. We used Ubuntu 16.04 LTS 64-bit operating system. 

We used 32 cores for the experiments. The system execution times to process the entire development set is $\approx$ 8 hours as this involves experimental optimization. However, for the evaluation set, the time required is almost identical to the baseline system ($\approx$ 1 hour). The additional time is just about 5 seconds per audio recording which includes i-vector extraction, domain-prediction and looktable search.

\section{Conclusion}
Our study is a step towards advancing the baseline diarization system with domain-dependent processing. Our system showed substantially reduced error rates as we optimized the clustering threshold and the dimensionality reduction parameters for each domain separately. The future work involves investigating advanced embedding extractors and exploring more domain-dependent processing, e.g., domain-dependent acoustic front-end, embedding extractor, re-segmentation, etc.

\section{Acknowledgements}
Experiments presented in this paper were partially carried out using the Grid'5000 testbed, supported by a scientific interest group hosted by Inria and including CNRS, RENATER and several Universities as well as other organizations (see \url{https://www.grid5000.fr}).

\bibliographystyle{IEEEtran}

\bibliography{mybib}

\end{document}